\documentclass[12pt]{article}
\usepackage{amssymb}
\usepackage{amsmath}

\newcommand{\be}{\begin{eqnarray}}
\newcommand{\ee}{\end{eqnarray}}
\newcommand{\non}{\nonumber}
\newcommand{\id}{\mathbb{I}}
\newcommand{\tr}{\mathop{\rm tr}\nolimits}

\begin{document}

\begin{titlepage}
\strut\hfill UMTG--281
\vspace{.5in}
\begin{center}

\LARGE Twisting singular solutions of Bethe's equations\\
\vspace{1in}
\large Rafael I. Nepomechie \footnote{nepomechie@physics.miami.edu}
and Chunguang Wang \footnote{c.wang22@umiami.edu}\\[0.8in]
\large Physics Department, P.O. Box 248046, University of Miami\\[0.2in]  
\large Coral Gables, FL 33124 USA\\

\end{center}

\vspace{.5in}

\begin{abstract}
    The Bethe equations for the periodic XXX and XXZ spin chains admit 
    singular solutions, for which the corresponding eigenvalues and eigenvectors are ill-defined.
    We use a twist regularization to derive conditions for such 
    singular solutions to be physical, in which case they correspond 
    to genuine eigenvalues and eigenvectors of the Hamiltonian.
\end{abstract}

\end{titlepage}

\setcounter{footnote}{0}

\section{Introduction}\label{sec:intro}

The periodic spin-1/2 isotropic Heisenberg (XXX) quantum spin chain, 
whose Hamiltonian is given by\footnote{We denote by $\vec \sigma = 
(\sigma^{x}\,, \sigma^{y}\,, \sigma^{z})$ the standard Pauli spin matrices
\be
\sigma^{x}=\left(\begin{array}{cc}
0 & 1 \\
1 & 0
\end{array}\right)\,, \quad
\sigma^{y}=\left(\begin{array}{cc}
0 & -i \\
i & 0
\end{array}\right)\,, \quad
\sigma^{z}=\left(\begin{array}{cc}
1 & 0 \\
0 & -1
\end{array}\right)\,, \non
\ee
which act on a two-dimensional complex vector space $V ={\cal C}^{2}$.
Moreover, $\sigma^{i}_{n}$ denotes an operator on the $N$-fold tensor product 
space $V^{\otimes N}$, which acts as $\sigma^{i}$ on the $n^{th}$ 
copy of $V$, and as the identity operator otherwise
\be
\sigma^{i}_{n} = \id \otimes \cdots \otimes  \id \otimes 
\stackrel{\stackrel{n}{\downarrow}}{\sigma^{i}} 
\otimes \id 
\cdots \otimes  \id \,, \non
\ee
where here $\id$ denotes the $2 \times 2$ identity matrix.
}
\be
H =  \frac{1}{4}\sum_{n=1}^{N}  \left(
\vec \sigma_{n} \cdot \vec \sigma_{n+1} - 1 \right)  \,, \qquad \vec
\sigma_{N+1} \equiv \vec \sigma_{1} \,,
\label{Hamiltonian}
\ee
is well known to be solvable by Bethe ansatz: both the eigenvectors
and the eigenvalues of the Hamiltonian can be expressed in
terms of solutions of the Bethe equations  \cite{Bethe:1931hc, Faddeev:1996iy}
\be
\left(\frac{\lambda_{j}+\frac{i}{2}}{\lambda_{j}-\frac{i}{2}}\right)^{N} 
 = \prod_{\scriptstyle{k \ne j}\atop \scriptstyle{k=1}}^{M}
\frac{\lambda_{j}-\lambda_{k}+i}{\lambda_{j}-\lambda_{k}-i} 
\,, \qquad j = 1, \ldots, M \,, \qquad M = 0, 1, \ldots, \frac{N}{2} \,.
\label{BE}
\ee
Indeed, the eigenvectors are given in terms of the ``Bethe roots'' 
$\{ \lambda_{j} \}$ by the Bethe vector 
\be
\prod_{j=1}^{M}B(\lambda_{j}) |0\rangle \,,
\label{Bethevector}
\ee
where $B(\lambda)$ is a certain creation operator\footnote{We remind 
the reader that the monodromy matrix is given by \cite{Faddeev:1996iy}
\be
T_{a}(\lambda) = L_{N a}(\lambda)  \cdots L_{1 a}(\lambda) = \left(
\begin{array}{cc}
    A(\lambda) & B(\lambda) \\
    C(\lambda) & D(\lambda) 
\end{array}    \right)  \,, \non 
\ee 
where the Lax operator is $L_{n a}(\lambda) = (\lambda - 
\frac{i}{2})\id_{n a} + i {\cal P}_{n a}$, and ${\cal P}$ is the permutation matrix 
on $V \otimes V$. The operator $B(\lambda)$ serves as a creation operator for 
constructing the eigenstates of $H$, and has the property 
$\left[B(\lambda)  \,, B(\lambda') \right] = 0$. The transfer matrix 
$t(\lambda)=\tr_{a} T_{a}(\lambda)=A(\lambda) + D(\lambda)$ satisfies $\left[t(\lambda)  \,, 
t(\lambda') \right] = 0$,
and therefore is the generator of commuting quantities $H_{n}=\frac{i}{2}\frac{d^{n}}{d\lambda^{n}}\log 
t(\lambda)\vert_{\lambda=\frac{i}{2}}$, with $H=H_{1}-\frac{N}{2}$.}  
and $|0\rangle$ is 
the state with all $N$ spins up; and the corresponding eigenvalues are given by
\be
E = -\frac{1}{2}\sum_{j=1}^{M}\frac{1}{\lambda_{j}^{2}+\frac{1}{4}} 
\label{energy}
\,.
\ee
It is also well known that the Bethe equations admit so-called
singular (or exceptional) solutions, for which the corresponding
eigenvectors and eigenvalues are ill-defined. (See e.g.
\cite{Avdeev:1985cx, Alcaraz:1987zr, Essler:1991wd, Siddharthan:1998, Noh:2000,
Beisert:2003xu, Beisert:2004hm, Hagemans:2007, Goetze:2010, 
Bazhanov:2010ts, Arutyunov:2012tx, Nepomechie:2013mua}.)   The
simplest example occurs for $M=2$ and any $N \ge 4$, namely $(\lambda_{1}, \lambda_{2}) 
= (i/2\,, - i/2)$.  To see that this is an exact solution, it is 
convenient to rewrite the Bethe equations (\ref{BE}) in polynomial form (see 
e.g. (\ref{BE2}) below). The corresponding energy (\ref{energy}) is evidently 
ill-defined, and the corresponding eigenvector (\ref{Bethevector}) can be shown to 
be null.

A general singular solution of the Bethe equations has the 
form
\be
\{ \frac{i}{2}\,, -\frac{i}{2}\,, \lambda_{3}\,, \ldots\,, \lambda_{M} \}\,,
\label{singular}
\ee
where $\lambda_{3}\,, \ldots\,, \lambda_{M}$ are distinct and not 
equal to $\pm i/2$. A solution that does not contain 
$\pm i/2$ is called regular.  Note that the order of the Bethe roots does not 
matter, since the Bethe equations (\ref{BE}) as well as the 
eigenvectors (\ref{Bethevector}) and eigenvalues (\ref{energy}) are 
invariant under any permutation of $\{\lambda_{1}\,, \ldots\,, 
\lambda_{M} \}$.

It is important to recognize that there are two main types of singular 
solutions:
{\bf physical} singular solutions (which correspond to genuine eigenvalues
and eigenvectors of the Hamiltonian), and {\bf unphysical} singular 
solutions (which do not correspond to eigenvalues and eigenvectors of 
the Hamiltonian). The simplest example of the former is $\pm i/2$ for 
$N$ even, while the simplest example of the latter is $\pm i/2$ for 
$N$ odd.

We have argued in \cite{Nepomechie:2013mua} that a general singular
solution (\ref{singular}) is physical if  $\lambda_{3},
\ldots , \lambda_{M}$ satisfy the following additional condition 
\be
\left[-\prod_{j=3}^{M}\left( \frac{\lambda_{j} + \frac{i}{2}}
{\lambda_{j} - \frac{i}{2}} \right)\right]^{N} =1 \,.
\label{consistency1}
\ee
For the case $M=2$, this condition reduces to the requirement 
(already noted above) that $N$ should be even.

This condition was used in \cite{Hao:2013jqa} to explicitly demonstrate the 
completeness of the solutions of Bethe's equations up to $N=14$. That 
is, the number of regular solutions plus the 
number of physical singular solutions (i.e., those singular solutions that satisfy 
(\ref{consistency1})) exactly coincides with the number needed to 
account for all $2^{N}$ eigenstates of the model. For further 
discussions of the completeness problem, see for example 
\cite{Bethe:1931hc, Faddeev:1996iy, Hagemans:2007,
Takahashi1971, Kirillov:1985, Tarasov:1995a, Tarasov:1995b, 
Langlands:1995, Langlands:1997, Fabricius:2000yx, Baxter:2001sx, 
Mukhin:2009}.

For the integrable spin-$s$ XXX chain, a generalization of 
(\ref{consistency1}) was derived and used to investigate completeness
in \cite{Hao:2013rza}. For related recent developments, see \cite{Kirillov:2014a,
Kirillov:2014b, Deguchi:2014fia}.

The derivation of the constraint (\ref{consistency1}) in \cite{Nepomechie:2013mua}
(and similarly of its spin-$s$ generalization in \cite{Hao:2013rza}) 
relies on regularizing the singular solution (\ref{singular}) by 
replacing the first two roots by
\be
\lambda_{1} = \frac{i}{2} + \epsilon + c\, \epsilon^{N}\,, \qquad 
\lambda_{2} = -\frac{i}{2} + \epsilon \,,
\label{regularized}
\ee
where $\epsilon$ is a small parameter, and $c$ is a constant that is still to 
be determined. This way of
regularizing a singular solution
was considered previously in \cite{Avdeev:1985cx, Beisert:2003xu, 
Beisert:2004hm, Hagemans:2007}.
Requiring that the corresponding Bethe vector (constructed as in 
(\ref{Bethevector}), except with a different normalization of the 
creation operators, namely $B(\lambda) \mapsto 
(\lambda+\frac{i}{2})^{-N} B(\lambda)$, which diverges at $\lambda=-\frac{i}{2}$) be an eigenvector of 
the transfer matrix in the limit $\epsilon 
\rightarrow 0$ gives rise to two equations for the constant $c$, 
whose consistency implies (\ref{consistency1}).

The regularization scheme (\ref{regularized}) may be rightly
criticized as being somewhat unphysical and ad-hoc.  Moreover, one can
worry that a different choice of regularization could
lead to a result different from (\ref{consistency1}). The primary
motivation for the present work was to see whether this constraint 
could be derived using a different, and more physical, regularization.

An alternative regularization is to introduce a small diagonal
twist angle $\beta$ in the boundary conditions (see e.g. \cite{Alcaraz:1987zr})
\be
\sigma^{x}_{N+1}&=&\cos \beta\, \sigma^{x}_{1} - \sin \beta\, 
\sigma^{y}_{1}\,,   \non \\
\sigma^{y}_{N+1}&=&\sin \beta\, \sigma^{x}_{1} + \cos \beta\, 
\sigma^{y}_{1}\,,   \non \\
\sigma^{z}_{N+1}&=&\sigma^{z}_{1} \,.
\label{twistBC}
\ee 
This boundary condition evidently breaks the $SU(2)$ symmetry down to 
$U(1)$, and reduces to periodic boundary conditions when $\beta = 0$. 

This way of regularizing a singular solution
was considered previously in \cite{Siddharthan:1998, Goetze:2010, 
Bazhanov:2010ts, Arutyunov:2012tx}. Moreover, such twists have been widely 
used in related contexts (see e.g. \cite{Bazhanov:2010ts, Tarasov:1995a, 
Tarasov:1995b, Langlands:1995, Kitanine:2004cp, Dorey:2007zx, Boos:2009} and references therein).
Like (\ref{regularized}), the twist regularization (\ref{twistBC}) involves 
introducing an additional parameter; however, the latter 
regularization is arguably more physical, since its parameter has a 
physical meaning.

We show here that the constraint (\ref{consistency1}) can indeed be 
derived (in fact, more easily) using the twist (\ref{twistBC}) as a 
regulator. The argument easily generalizes to the case of arbitrary 
spin $s$, and also to the XXZ case.

\section{XXX}

For the spin-1/2 XXX spin chain with twisted boundary conditions 
(\ref{twistBC}), the Bethe equations are given by
\be
\left(\frac{\lambda_{j}+\frac{i}{2}}{\lambda_{j}-\frac{i}{2}}\right)^{N} 
 = e^{-i\beta}
\prod_{\scriptstyle{k \ne j}\atop \scriptstyle{k=1}}^{M}
\frac{\lambda_{j}-\lambda_{k}+i}{\lambda_{j}-\lambda_{k}-i} 
\,, \qquad j = 1, \ldots, M \,, 
\label{BE1}
\ee
which can be rewritten in polynomial form as 
\be
\left(\lambda_{j}+\frac{i}{2}\right)^{N} 
\prod_{\scriptstyle{k \ne j}\atop \scriptstyle{k=1}}^{M}
(\lambda_{j}-\lambda_{k}-i) = e^{-i\beta}
\left(\lambda_{j}-\frac{i}{2}\right)^{N} 
 \prod_{\scriptstyle{k \ne j}\atop \scriptstyle{k=1}}^{M}
(\lambda_{j}-\lambda_{k}+i) \,, \qquad j = 1, \ldots, M \,.
\label{BE2}
\ee

We assume that, for small $\beta$, the roots
$\pm i/2$ of a physical singular solution (\ref{singular}) acquire corrections of 
order $\beta$,\footnote{The twisted equations (\ref{BE2}) evidently still 
admit solutions with $\pm i/2$ (i.e., without any $\beta$-dependent 
corrections). However, such singular solutions are unphysical.}
\be
\lambda_{1} &= \frac{i}{2} + c_{1} \beta + O(\beta^{2}) \,, \non\\
\lambda_{2} &= -\frac{i}{2} + c_{2} \beta + O(\beta^{2})  \,,
\label{pmi2}
\ee
where $c_{1}$ and $c_{2}$ are some constants (independent of $\beta$).
The Bethe equations (\ref{BE2}) for $\lambda_{1}$ and $\lambda_{2}$ are
\be
\left(\lambda_{1}+\frac{i}{2}\right)^{N}(\lambda_{1}-\lambda_{2}-i)\prod_{k=3}^{M}(\lambda_{1}-\lambda_{k}-i) = e^{-i\beta}
\left(\lambda_{1}-\frac{i}{2}\right)^{N}(\lambda_{1}-\lambda_{2}+i) 
\prod_{k=3}^{M}(\lambda_{1}-\lambda_{k}+i) \,, \non \\
\left(\lambda_{2}+\frac{i}{2}\right)^{N}(\lambda_{2}-\lambda_{1}-i)
\prod_{k=3}^{M}(\lambda_{2}-\lambda_{k}-i) = e^{-i\beta}
\left(\lambda_{2}-\frac{i}{2}\right)^{N}(\lambda_{2}-\lambda_{1}+i) 
\prod_{k=3}^{M}(\lambda_{2}-\lambda_{k}+i) \,.\non 
\ee
Substituting (\ref{pmi2}), one can see that these equations are 
satisfied to first order in $\beta$ provided that
\be
c_{1} = c_{2} \,.
\label{c1c2}
\ee 
Forming the product of all $M$ Bethe equations (\ref{BE1}), we obtain
\be
\left(\frac{\lambda_{1}+\frac{i}{2}}{\lambda_{1}-\frac{i}{2}}
\frac{\lambda_{2}+\frac{i}{2}}{\lambda_{2}-\frac{i}{2}}
\prod_{j=3}^{M}\frac{\lambda_{j}+\frac{i}{2}}{\lambda_{j}-\frac{i}{2}}\right)^{N} 
 = e^{-i M\beta} 
 \,. \label{prod}
\ee
Substituting (\ref{pmi2}) and (\ref{c1c2}) into (\ref{prod}) and taking the limit 
$\beta \rightarrow 0$, we arrive at the constraint (\ref{consistency1})
\be
\left[-\prod_{j=3}^{M}\left( \frac{\lambda_{j} + \frac{i}{2}}
{\lambda_{j} - \frac{i}{2}} \right)\right]^{N} =1 \,.
\label{consistency}
\ee
This concludes our argument for the spin-1/2 
XXX case. Of course,  $\{\lambda_{3}, \ldots , \lambda_{M} \}$ must  also obey
\be
\left(  \frac{\lambda_{j} + \frac{i}{2}}
{\lambda_{j} - \frac{i}{2}} \right)^{N-1} 
\left(\frac{\lambda_{j} - \frac{3i}{2}}
{\lambda_{j} + \frac{3i}{2}}\right)
=  \prod_{\scriptstyle{k \ne j}\atop \scriptstyle{k=3}}^{M}
\frac{\lambda_{j} - \lambda_{k} + i} 
{\lambda_{j} - \lambda_{k} - i}
\,, \qquad j = 3 \,, \cdots \,, M \,,
\label{BAE3}
\ee
which follow from the Bethe equations (\ref{BE1}) with $j=3, \ldots, 
M$ after substituting (\ref{pmi2}) and taking $\beta \rightarrow 0$.

The constraint (\ref{consistency}) can also be derived in a similar 
way using the original regularization (\ref{regularized}) simply by substituting 
into (\ref{prod}) (with $\beta=0$) and taking the limit $\epsilon 
\rightarrow 0$. This argument (which was overlooked in 
\cite{Nepomechie:2013mua}) evidently does not require the $ c 
\epsilon^{N}$ term in  (\ref{regularized}). However, this $ c 
\epsilon^{N}$ term is needed to construct the correct eigenvector.

In order to construct the eigenvector corresponding to a physical 
singular solution using the twist regularization, we expect (based on 
\cite{Nepomechie:2013mua}) that it is necessary 
to determine the corrections of the singular solution up to order $\beta^{N}$, to 
renormalize the Bethe vector (\ref{Bethevector}) by the factor  $1/\beta^{N}$, and then take 
the limit $\beta \rightarrow 0$.
The required corrections of the singular solution can be obtained (for given
explicit values $\{\lambda_{j}^{(0)}\}$ of $\{\lambda_{3}, \ldots , \lambda_{M} \}$ that satisfy
(\ref{consistency}) and (\ref{BAE3}))
by assuming that all the Bethe 
roots can be expanded in powers of $\beta$,
\be
\lambda_{j} = \lambda_{j}^{(0)} + \sum_{l=1}^{N} c_{j}^{(l)} 
\beta^{l} +  O(\beta^{N+1})
 \,, \qquad j = 1\,, \ldots \,, M\,, 
\ee
and solving the Bethe equations (\ref{BE2}), (\ref{prod})  for the coefficients 
$c_{j}^{(l)}$. For example, for the simplest case $(N, M) = (4, 2)$, we find in 
this way
\be
\lambda_{1} &= \frac{i}{2} + \frac{\beta}{4} - \frac{\beta^{3}}{96} + 
\frac{i \beta^{4}}{256} + O(\beta^{5}) \,, \non \\
\lambda_{2} &= -\frac{i}{2} + \frac{\beta}{4} - \frac{\beta^{3}}{96} - 
\frac{i \beta^{4}}{256} + O(\beta^{5}) \,.
\ee
Moreover, we have verified by explicit computation that the vector 
\be
\lim_{\beta \rightarrow 0} \frac{1}{\beta^{4}}B(\lambda_{1})\, B(\lambda_{2}) 
|0\rangle 
\ee 
is indeed proportional to the correct eigenvector \cite{Avdeev:1985cx, Essler:1991wd}
$\sum_{k=1}^{4}(-1)^{k} S_{k}^{-} S_{k+1}^{-} |0\rangle $.

\subsection{Spin $s$}

Similar arguments can be applied to the integrable spin-$s$ XXX chain 
with twisted boundary conditions, 
for arbitrary spin $s = \frac{1}{2}, 1, \frac{3}{2}, \ldots$. The 
Bethe equations are given by
\be
\left(\frac{\lambda_{j}+i s}{\lambda_{j}-i s}\right)^{N} 
 = e^{-i\beta}
\prod_{\scriptstyle{k \ne j}\atop \scriptstyle{k=1}}^{M}
\frac{\lambda_{j}-\lambda_{k}+i}{\lambda_{j}-\lambda_{k}-i} 
\,, \qquad j = 1, \ldots, M \,, 
\label{BEs}
\ee
When $\beta=0$, these equations have singular solutions of the form \cite{Hao:2013rza}
\be
\{ i s\,, i(s-1)\,, \ldots \,, -i(s-1)\,, -i s \,, \lambda_{2s+2}\,, 
\ldots, \lambda_{M}\} \,,
\label{singulars}
\ee
where all the roots are assumed to be distinct.
That is, a singular solution contains an exact string of length $2s+1$ centered at the origin.

We assume that, for small $\beta$, the roots $\{ i s\,, i(s-1)\,,
\ldots \,, -i(s-1)\,, -i s \}$ of a physical singular solution acquire
corrections of order $\beta$,
\be
\lambda_{k} &= i (s+1-k) + c_{k} \beta + O(\beta^{2}) \,, \qquad k = 
1\,,2 \,, \ldots \,, 2s+1 \,,
\label{singulars2}
\ee
where $\{ c_{k} \}$ are some constants.
Substituting (\ref{singulars2}) into the first $2s+1$ Bethe equations
(i.e., Eq.  (\ref{BEs}) for $j = 1\,, \ldots \,, 2s+1$), we see that these equations are 
satisfied to first order in $\beta$ provided that all the $c_{k}$'s are 
equal,
\be
c_{1} = c_{2} = \ldots = c_{2s+1}\,.
\label{c1c2s}
\ee 
The product of all $M$ Bethe equations (\ref{BEs}) gives
\be
\left(\frac{\lambda_{1}+is}{\lambda_{1}-is}
\frac{\lambda_{2}+is}{\lambda_{2}-is}
\cdots
\frac{\lambda_{2s+1}+is}{\lambda_{2s+1}-is}
\prod_{j=2s+2}^{M}\frac{\lambda_{j}+is}{\lambda_{j}-is}\right)^{N} 
 = e^{-i M\beta} 
\,. \label{prods}
\ee
Substituting (\ref{singulars2}) and (\ref{c1c2s}) into (\ref{prods}) and taking the limit 
$\beta \rightarrow 0$, we obtain the constraint
\be
\left[(-1)^{2s}\prod_{j=2s+2}^{M}\left( \frac{\lambda_{j} + i s}
{\lambda_{j} - is } \right)\right]^{N} =1 \,.
\label{consistency2} 
\ee
This necessary condition for the singular solution (\ref{singulars}) 
to be physical, which is evidently a generalization of the $s=1/2$
result (\ref{consistency}), was first obtained in \cite{Hao:2013rza} 
using instead a generalization of the regularization (\ref{regularized}).
Of course, $\{\lambda_{2s+2}, \ldots , \lambda_{M} \}$ must  also obey
\be
\left( \frac{\lambda_{j} + is}{\lambda_{j} - is}\right)^{N-1} 
\left(\frac{\lambda_{j} - i(s+1)}{\lambda_{j} + i(s+1)} \right) = 
\prod_{\scriptstyle{k \ne j}\atop \scriptstyle{k=2s+2}}^M 
\frac{\lambda_{j} -\lambda_{k} +i}{\lambda_{j} -\lambda_{k} -i}
\,, \qquad j = 2s+2 \,, \ldots \,, M \,,
\ee 
which follow from the Bethe equations (\ref{BEs}) with $j=3, \ldots, 
M$ after substituting (\ref{singulars2}) and taking $\beta \rightarrow 0$.

\section{XXZ}

For the spin-1/2 XXZ spin chain with twisted boundary conditions, the Bethe equations are given by
\be
\left(\frac{\sinh(\lambda_{j}+\frac{\eta}{2})}{\sinh(\lambda_{j}-\frac{\eta}{2})}\right)^{N} 
 = e^{-i\beta}
\prod_{\scriptstyle{k \ne j}\atop \scriptstyle{k=1}}^{M}
\frac{\sinh(\lambda_{j}-\lambda_{k}+\eta)}{\sinh(\lambda_{j}-\lambda_{k}-\eta)} 
\,, \qquad j = 1, \ldots, M \,, 
\label{BEXXZ}
\ee
where $\eta$ is the anisotropy parameter, which we assume has a generic 
value (i.e., $q=e^{\eta}$ is not a root of unity). When $\beta=0$, these 
equations have singular solutions of the form   
\be
\{ \frac{\eta}{2}, -\frac{\eta}{2}, \lambda_{3}, \ldots , \lambda_{M} 
\}\,.
\label{singularXXZ}
\ee
Repeating the same steps of our argument for the isotropic case, we 
conclude that a physical singular solution must satisfy the constraint
\be
\left[-\prod_{j=3}^{M}
\frac{\sinh(\lambda_{j}+\frac{\eta}{2})}{\sinh(\lambda_{j}-\frac{\eta}{2})}
\right]^{N} =1 \,,
\label{consistencyXXZ}
\ee
as well as 
\be
\left(  \frac{\sinh(\lambda_{j} + \frac{\eta}{2})}
{\sinh(\lambda_{j} - \frac{\eta}{2})} \right)^{N-1} 
\frac{\sinh(\lambda_{j} - \frac{3\eta}{2})}
{\sinh(\lambda_{j} + \frac{3\eta}{2})}
=  \prod_{\scriptstyle{k \ne j}\atop \scriptstyle{k=3}}^{M}
\frac{\sinh(\lambda_{j} - \lambda_{k} + \eta)} 
{\sinh(\lambda_{j} - \lambda_{k} - \eta)}
\,, \qquad j = 3 \,, \cdots \,, M \,.
\ee

Similarly, for the spin-s XXZ spin chain with twisted boundary conditions, 
the Bethe equations are given by
\be
\left(\frac{\sinh(\lambda_{j}+ s \eta)}{\sinh(\lambda_{j}-s \eta)}\right)^{N} 
 = e^{-i\beta}
\prod_{\scriptstyle{k \ne j}\atop \scriptstyle{k=1}}^{M}
\frac{\sinh(\lambda_{j}-\lambda_{k}+\eta)}{\sinh(\lambda_{j}-\lambda_{k}-\eta)} 
\,, \qquad j = 1, \ldots, M \,.
\label{BEXXZs}
\ee
When $\beta=0$, these equations have singular solutions of the form   
\be
\{ s \eta\,, (s-1)\eta\,, \ldots \,, -(s-1)\eta\,, - s \eta \,, \lambda_{2s+2}\,, 
\ldots, \lambda_{M}\} \,,
\label{singularXXZs}
\ee
where again all the roots are assumed to be distinct.
A physical singular solution of this form must satisfy the constraint
\be
\left[(-1)^{2s}\prod_{j=2s+2}^{M}
\frac{\sinh(\lambda_{j}+s \eta)}{\sinh(\lambda_{j}-s \eta)}
\right]^{N} =1 \,,
\label{consistencyXXZs}
\ee
as well as 
\be
\left( \frac{\sinh(\lambda_{j} + s \eta)}{\sinh(\lambda_{j} - s \eta)}\right)^{N-1} 
\frac{\sinh(\lambda_{j} - (s+1) \eta)}{\sinh(\lambda_{j} + (s+1) \eta)}
= \prod_{\scriptstyle{k \ne j}\atop \scriptstyle{k=2s+2}}^M 
\frac{\sinh(\lambda_{j} -\lambda_{k} +\eta)}{\sinh(\lambda_{j} 
-\lambda_{k} -\eta)}
\,, \quad j = 2s+2 \,, \ldots \,, M \,.
\ee 

The constraint (\ref{consistencyXXZ}) and its generalization (\ref{consistencyXXZs}), 
which heretofore had not been written down, can also be 
straightforwardly derived using the alternative regularization 
(\ref{regularized}) following \cite{Nepomechie:2013mua} and \cite{Hao:2013rza}.

\section{Conclusion}

We have argued that a twist regularization can be used to derive the
constraints (\ref{consistency}), (\ref{consistency2}),
(\ref{consistencyXXZ}), (\ref{consistencyXXZs}) for singular solutions
of the periodic XXX and XXZ spin chains to be physical.  The fact that
these constraints can be derived using two different regularizations
suggests that they are independent of the choice of regularization.
Indeed, the fact that these constraints appear already at first order
in the regulator (instead of order $N$, as suggested by the original
derivations \cite{Nepomechie:2013mua, Hao:2013rza}) implies that they
are robust.

Although the arguments presented here demonstrate only that these 
conditions are necessary, the arguments in \cite{Nepomechie:2013mua} 
and \cite{Hao:2013rza} imply that these conditions are also sufficient for 
singular solutions to be physical. This conclusion is also supported 
by numerical evidence \cite{Hao:2013jqa, 
Hao:2013rza}. The latter references also show that most of the solutions 
of the Bethe equations are unphysical singular solutions; hence, it 
is all the more important to have simple criteria for picking out from among 
the many singular solutions the few that are physical.

As noted in \cite{Avdeev:1985cx, Hao:2013rza}, the Bethe equations for
chains with $s>1/2$ can also have singular solutions with
repeated roots that are physical.  We expect that the twist
regularization considered here can also be used to derive conditions
for such ``strange'' singular solutions to be physical.

\section*{Acknowledgments}
We are grateful to Nikita Slavnov for sharing with us his unpublished 
notes on the effect of a twist on the singular solution $\pm \eta/2$,
for valuable correspondence, and for his comments on a draft.
We also thank Jean-Sebastien Caux for helpful discussions; and 
many colleagues, including Patrick Dorey, Frank  
G\"ohmann, Ivan Kostov and Matthias Staudacher, for their gentle 
prodding to consider the twist regularization.
The work of RN was supported in part by the National Science
Foundation under Grant PHY-1212337, and by a Cooper fellowship.


\providecommand{\href}[2]{#2}\begingroup\raggedright\endgroup

\end{document}